\documentclass
[superscriptaddress,secnumarabic,amssymb,amsmath,nobibnotes,aps,prd,showkeys,showpacs,nofootinbib,onecolumn,12pt]{revtex4}%
\usepackage{graphicx}
\usepackage{epsf}
\usepackage{bm}
\usepackage{amsmath}
\usepackage{amsfonts}
\usepackage{amssymb}%
\providecommand{\U}[1]{\protect\rule{.1in}{.1in}}

\newcommand{\be}{\begin{equation}}
\newcommand{\ee}{\end{equation}}

\newcommand{\mincir}{\raise
-3.truept\hbox{\rlap{\hbox{$\sim$}}\raise4.truept\hbox{$<$}\ }}
\newcommand{\magcir}{\raise
-3.truept\hbox{\rlap{\hbox{$\sim$}}\raise4.truept\hbox{$>$}\ }}

\begin{document}
\title{Dynamics in interacting scalar-torsion theory}
\author{Andronikos Paliathanasis}
\email{anpaliat@phys.uoa.gr}
\affiliation{Institute of Systems Science, Durban University of Technology, Durban 4000,
South Africa }
\affiliation{Instituto de Ciencias F\'{\i}sicas y Matem\'{a}ticas, Universidad Austral de
Chile, Valdivia 5090000, Chile}

\begin{abstract}
In a spatially flat \ Friedmann--Lema\^{\i}tre--Robertson--Walker background
space we consider a scalar-torsion gravitational model which has similar
properties with the dilaton theory. This teleparallel model is invariant under
a discrete transformation similar to the Gasperini-Veneziano duality
transformation. Moreover, in the gravitational Action integral we introduce
the Lagrangian function of a pressureless fluid source which is coupled to the
teleparallel dilaton field. This specific gravitational theory with
interaction in the dark sector of the universe is investigated by using
methods of the dynamical system analysis. We calculate that the theory
provides various areas of special interest for the evolution of the
cosmological history. Inflationary scaling solutions and the de Sitter
universe is recovered. Furthermore, we calculate that there exist an attractor
which provides a stable solution where the two fluid components, the scalar
field and the pressureless matter, contribute in the cosmological fluid. This
solution is of special interest because it can describe the present epoch.
Finally, the qualitative evolution of the cosmographic parameters is discussed.

\end{abstract}
\date{\today}
\maketitle

\section{Introduction}

\label{sec1}

The detailed analysis of the recent cosmological data indicates that General
Relativity may need to be modified in order to describe the observations, for
a recent review we refer the reader to \cite{sp1}. The late-time cosmic
acceleration has been attributed to a fluid, the so-called dark energy, with
has negative pressure and anti-gravity effects \cite{dr}. In order to explain
the anti-gravitational effects, cosmologists have proposed the modification of
the Einstein-Hiblert Action by using geometric invariants \cite{cl1}. In this
direction new geometrodynamical terms are introduced in the field equations
which can drive the dynamics in order to explain, with a geometric approach,
the observational phenomena \cite{cap1,odi}.

Teleparallelism \cite{ein28,Hayashi79} includes a class of modify theories of
gravity which have been widely studied the last years
\cite{fg5,ft5,ft6,ft0,ft1,ft2,ft3,ft4}. The fundamental invariant of
teleparallelism is the torsion scalar of the antisymmetric connection which
play the role important role, instead of the Levi-Civita connection in General
Relativity. In previous studies it has been discussed that teleparallel
gravity my violates the Lorentz symmetry \cite{li}, while Lorentz violation
has not been observed yet, it is common in various subjects of quantum gravity
\cite{dav1}. However, recent studies has shown that Lorentz symmetry can be
preserved in teleparallelism, see for instance \ \cite{lv1,lv2}. Specifically,
in \cite{lv1} it was found that the introduction of a scalar field in the
gravitational Action Integral of teleparallelism preserves the Lorentz
symmetry. For a recent review on teleparallelism we refer the reader in
\cite{rev1}. In the literature, teleparallel cosmology has been widely
studied. For instance, in $f\left(  T\right)  $ teleparallel theory the
cosmological perturbations were investigated in \cite{st1,st2}, while in
\cite{st2} it was found that $f\left(  T\right)  $ theory can mimicking
dynamical dark energy models. The mechanism which describe the Higgs inflation
era in scalar torsion theory was studied in \cite{st3}. An extension of the
scalar torsion theories with the introduction of the boundary term was
introduced in \cite{st4}. For other recent studies on teleparallelism we refer
the reader in \cite{st5,st6,st7,st8,st9} and references therein.

In this study we focus on the scalar-torsion or teleparallel dark energy
models \cite{te1,te2,te3} which can be seen as the analogue of the
scalar-tensor theories in teleparallelism. In scalar-torsion theory an scalar
field is introduced in the gravitational Action Integral which is nonminimally
coupled to the fundamental scalar of teleparallelism, the torsion scalar.
There are various studies in the literature which indicate that the
scalar-torsion theories can explain the recent cosmological observations
\cite{te4,te5}. In the following we consider the existence of a matter source
with zero pressure in the gravitation Action which interacts with the scalar
field \cite{in1,in2,in3,in4}. In our analysis the interaction between the
scalar field and the pressureless fluid are inspired by the interaction
provided in the Weyl integrable theory \cite{win1,win2}. The plan of the paper
is as follows.

In Section \ref{sec2} we introduce the model of our consideration which is the
teleparallel dilaton model coupled to a pressureless fluid source. This model
belongs to the family of scalar-torsion theories, the field equations are
derived. Section \ref{sec3} includes the new material of this study. We
performed a detailed study of the asymptotic dynamics for the gravitational
field equations for the model of our consideration. We determine the
stationary points and we investigate their stability as we discuss the
physical properties of the exact solutions described by the stationary points.
This analysis provide important information about the cosmological viability
of the proposed theory. It is clear that our model can explain the mayor eras
of the cosmological evolution and it can be used as a dark energy candidate.
Furthermore, in Section \ref{sec4} we discuss the evolution of the
cosmographic parameters as they are provided by our model. Finally, in Section
\ref{sec5} we summarize our results and we draw our conclusions.

\section{Teleparallel dilaton model}

\label{sec2}

The gravitational model of our consideration is an extension of the
teleparallel dilaton theory known as scalar-torsion theory where the
gravitational Action Integral is defined
\begin{equation}
S=\frac{1}{16\pi G}\int d^{4}xe\left[  e^{-\frac{\phi}{2}}\left(
T+\frac{\omega}{2}\phi_{;\mu}\phi^{;\mu}+V\left(  \phi\right)  +L_{m}\right)
\right]  ,\label{td.01}%
\end{equation}
in which $T$ is the torsion scalar of the antisymmetric curvatureless
connection, $\phi\left(  x^{k}\right)  $ is a scalar field with potential
function $V\left(  \phi\right)  ,~\omega$ is a nonzero constant, $L_{m}$ is
the Lagrangian function for the additional matter source and $e$ is the
determinant of the vierbein fields. Action Integral (\ref{td.01}) belongs to
the family of gravitational models known as teleparallel dark energy models,
or scalar torsion models \cite{te1,te2,ss2,ss3}. Scalar-torsion models can be
seen as the analogue of scalar-tensor models in\ teleparallelism, in which
instead of the Ricci scalar the torsion scalar $T$ is used for the definition
of the Action Integral. The gravitational field equations are of second-order,
however, under conformal transformations the scalar-torsion theories are now
equivalent with the quintessence model \cite{ss2}. Under a conformal
transformation the scalar-torsion Action Integral is equivalent with that of a
modified higher-order theory known as $f\left(  T,B\right)  $ where $B$ is the
boundary term differs the torsion $T$ and the Ricci scalar \cite{ss2}. That
makes the scalar-torsion and scalar-tensor theories to be totally different
which means that there is not any conformal transformation which may connect
the solutions of the two theories. In this study, we assume that the matter
source and the scalar field are interacting, that is, from (\ref{td.01}) the
mixed term $e^{-\frac{\phi}{2}}L_{m}$ exists.

In the case of a spatially flat Friedmann--Lema\^{\i}tre--Robertson--Walker
metric (FLRW) background space%
\begin{equation}
ds^{2}=-dt^{2}+a^{2}(t)(dx^{2}+dy^{2}+dz^{2}).\label{td.02}%
\end{equation}
the gravitational field equations for the (\ref{td.01}) in the case of vacuum
are invariant under a discrete transformation with origin the $O\left(
d,d\right)  $ symmetry \cite{odd}. The resulting discrete transformation is a
generalization of the Gasperini-Veneziano scale-factor duality transformation
for the dilaton field in scalar tensor theory \cite{ss1}. The existence of
this discrete transformation it is important for the study of the pre-big bang
period of the universe as it is described by string cosmology. However, when
parameter $\omega$ is small, the Gasperini-Veneziano transformation is
recovered. However, in general the pre-big bang period for the teleparallel
dilaton theory differs from that of sting cosmology by a term provided by the
nonzero constant $\omega$.

In this study we assume the scalar field $\phi$ interact with the matter
source. For the latter, we assume that of a pressureless fluid source, dust
fluid, with energy density $\rho_{m}$, known as dark matter. Models with
interaction in the dark sector of the universe has drawn the attention of
cosmologists the last decade \cite{int1,int2,int3,int4,int5}. Indeed, there
are various theoretical studies which shows that such models are viable, while
from the analysis of the cosmological data it seems that the interacting
models are supported by the observations \cite{int7,int8,int9,int10}. The
Action Integral (\ref{td.01}) can be seen as the teleparallel extension of
some scalar tensor models coupled to dark matter \cite{win1,win2}.

For the background space (\ref{td.02}) it follows $T=6H^{2}$, $H=\frac{\dot
{a}}{a}$, hence, from the Action integral (\ref{td.01}) and for a dust fluid
for the matter source we derive the modified gravitational field equations
\begin{equation}
e^{-\frac{\phi}{2}}\left(  6H^{2}-\left(  \frac{\omega}{2}\dot{\phi}%
^{2}+V\left(  \phi\right)  \right)  -\rho_{m}\right)  =0~,\label{td.03}%
\end{equation}%
\begin{equation}
\left(  2\dot{H}+3H^{2}\right)  +\frac{1}{2}\left(  \frac{\omega}{2}\dot{\phi
}^{2}-V\left(  \phi\right)  \right)  -H\dot{\phi}=0~,\label{td.04}%
\end{equation}
while for the matter source and the scalar field it follows%
\begin{equation}
\dot{\rho}_{m}+\left(  3H-\dot{\phi}\right)  \rho_{m}=0~,\label{td.05}%
\end{equation}%
\begin{equation}
\omega\left(  \dot{\phi}^{2}-2\ddot{\phi}-6H\dot{\phi}\right)  +2\left(
V-V_{,\phi}\right)  =0.\label{td.06}%
\end{equation}

In the case where $\rho_{m}=0~$and $V\left(  \phi\right)  =V_{0}$, the
discrete transformation which keeps invariant the field equations is
\begin{equation}
a\rightarrow a^{p_{1}}e^{p_{2}\phi}~,~\phi\rightarrow p_{3}\ln a+p_{4}%
\phi\label{td.06a}%
\end{equation}
in which $p_{1}=-\frac{1+3\omega}{1-3\omega}~$, $p_{2}=\frac{\omega}%
{1-3\omega}~$, $p_{3}=-\frac{12}{1-3\omega}$ and $p_{4}=\frac{1+3\omega
}{1-3\omega}$ when $\omega\neq\frac{1}{3}$. However, in the presence of the
matter source the field equations do not remain invariant under the action of
the discrete transformation. Moreover, when $\omega$ is near to zero the
discrete transformation (\ref{td.06a}) becomes the Gasperini-Veneziano duality
transformation \cite{ss1}. The discrete transformation (\ref{td.06a}) follows
from the presence of the $O\left(  d,d\right)  $ symmetry for the Action
Integral (\ref{td.01}). In addition, the existence of the $O\left(
d,d\right)  $ symmetry indicates the presence of a conservation law for the
classical field equations which can be used in order to integrate and write
the analytic solution in closed-form expression. Finally, the $O\left(
d,d\right)  $ symmetry is preserved and in the quantization process of the
theory. Hence, a quantum operator related to the $O\left(  d,d\right)  $
symmetry is determined which helps us to solve the Wheeler-DeWitt equation
\cite{odd}.

Additionally, from equation (\ref{td.03}) we define the energy density for the
scalar field $\rho_{\phi}=\left(  \frac{\omega}{2}\dot{\phi}^{2}+V\left(
\phi\right)  \right)  $, thus in order $\rho_{\phi}\geq0$ and do not have
ghost terms we assume $\omega>0$. The pressure component $p_{\phi}$ of the
scalar field from equation (\ref{td.04})~it is defined as $p_{\phi}=\frac
{1}{2}\left(  \frac{\omega}{2}\dot{\phi}^{2}-V\left(  \phi\right)  \right)
-H\dot{\phi}$.

In the following Section we study the general evolution for the cosmological
history as it is provided by the dynamical system (\ref{td.03})-(\ref{td.06}).

\section{Asymptotic dynamics}

\label{sec3}

We define the new dimensionless variables%
\begin{equation}
x=\sqrt{\frac{\omega}{12}}\frac{\dot{\phi}}{H}~,~y^{2}=\frac{V}{6H^{2}%
}~,~\Omega_{m}=\frac{\rho_{m}}{6H^{2}}~,~\lambda=\frac{V_{,\phi}}%
{V}~\label{td.07}%
\end{equation}
where the field equations are written as the following algebraic-differential
system%
\begin{equation}
\Omega_{m}=1-x^{2}-y^{2}\label{td.08}%
\end{equation}%
\begin{equation}
\frac{dx}{d\tau}=\frac{3}{2}x\left(  x^{2}-y^{2}-1\right)  +\sqrt{\frac
{3}{\omega}}y^{2}\left(  1-\lambda\right)  ~,\label{td.09}%
\end{equation}%
\begin{equation}
\frac{dy}{d\tau}=y\left(  \frac{3}{2}\left(  1+x^{2}-y^{2}\right)
-\sqrt{\frac{3}{\omega}}x\left(  1-\lambda\right)  \right)  ~,~\label{td.10}%
\end{equation}%
\begin{equation}
\frac{d\lambda}{d\tau}=2\sqrt{3}x\lambda^{2}\left(  \Gamma\left(
\lambda\right)  -1\right)  ~,~\Gamma\left(  \lambda\right)  =\frac
{V_{,\phi\phi}V}{\left(  V_{,\phi}\right)  ^{2}}~,\label{td.12}%
\end{equation}
in which the new independent variable is~$dt=Hd\tau$.

In addition, in the new variables the equation of state parameter for the
effective fluid $w_{eff}=-1-\frac{2}{3}\frac{\dot{H}}{H^{2}}$ reads%
\begin{equation}
w_{eff}\left(  x,y,\eta,\lambda\right)  =x^{2}-y^{2}-\frac{2}{\sqrt{3\omega}%
}x.\label{td.13}%
\end{equation}

As far as the scalar field potential is concerned, we consider the exponential
potential
\begin{equation}
V\left(  \phi\right)  =V_{0}e^{\lambda_{0}\phi}.
\end{equation}
For the exponential potential we infer that $\lambda=\lambda_{0}$ is a
constant and $\Gamma\left(  \lambda\right)  =1$ \ such that the rhs of
equation (\ref{td.12}) is always zero. Thus, for the exponential potential the
dimension of the dynamical system is two. Moreover, in order the matter source
to be physically accepted, it follows that $\Omega_{m}$ is constraint as,
$0\leq\Omega_{m}\leq1$, from (\ref{td.08}) it follows that the variables
$\left(  x,y\right)  $ are constrained similarly as $x^{2}+y^{2}\leq1$.
Furthermore, we observe that the equations are invariant under the change of
variables $y\rightarrow-y$, which means that without loss of generality we can
work on the branch $y\geq0$.

\subsection{Stationary points for the exponential potential}%

\begin{table}[tbp] \centering
\caption{Stationary points of the field equations and their physical
properties for the exponential potential.}%
\begin{tabular}
[c]{cccccc}\hline\hline
\textbf{Point} & $\left(  \mathbf{x,y}\right)  $ & \textbf{Existence} &
$\mathbf{\Omega}_{m}$ & $\mathbf{w}_{eff}$ & $\mathbf{w}_{eff}<-\frac{1}{3}%
$\\\hline
$A_{1}$ & $\left(  0,0\right)  $ & Always & $1$ & $0$ & No,~dust solution\\
$A_{2}^{\pm}$ & $\left(  1,0\right)  $ & Always & $0$ & $1\mp\frac{2}%
{\sqrt{3\omega}}$ & $A_{2}^{+}$ for $\omega>\frac{4}{3}$\\
$A_{3}$ & $\left(  \pm\frac{1-\lambda}{\sqrt{3\omega}},\sqrt{1-\frac{\left(
1-\lambda\right)  ^{2}}{3\omega}}\right)  $ & $\omega>\frac{1}{3}\left(
1-\lambda\right)  ^{2}$ & $0$ & $-1+\frac{2\lambda\left(  \lambda-1\right)
}{3\omega}$ & Yes under conditions\\
$A_{4}$ & $\left(  \frac{\sqrt{3\omega}}{2\left(  1-\lambda\right)  }%
,\sqrt{\frac{3\omega}{4\left(  1-\lambda\right)  ^{2}}}\right)  $ &
$\lambda\neq1~$ & $1-\frac{3\omega}{2\left(  1-\lambda\right)  ^{2}}$ &
$\frac{1}{\lambda-1} $ & Yes under conditions\\
&  & $\omega<\frac{2}{3}\left(  1-\lambda\right)  ^{2}$ &  &  & \\\hline\hline
\end{tabular}
\label{tab1}%
\end{table}%
%

\begin{table}[tbp] \centering
\caption{Stationary points of the field equations and their stability
properties for the exponential potential.}%
\begin{tabular}
[c]{ccc}\hline\hline
\textbf{Point} & \textbf{Eigenvalues} & \textbf{Stability}\\\hline
$A_{1}$ & $-\frac{3}{2}~,~\frac{3}{2}$ & Saddle\\
$A_{2}^{\pm}$ & $3\pm\sqrt{\frac{3}{\omega}}\left(  \lambda-1\right)  ~,~3$ &
$A_{2}^{+}$ saddle $\lambda<1$ and $\omega<\frac{1}{3}\left(  1-\lambda
\right)  ^{2}$\\
&  & $A_{2}^{-}$ saddle $\lambda>1$ and $\omega<\frac{1}{3}\left(
1-\lambda\right)  ^{2}$\\
$A_{3}$ & $-3+\frac{2}{\omega}\left(  1-\lambda\right)  ^{2},~-3+\frac{\left(
1-\lambda\right)  ^{2}}{\omega}$ & Stable for $\omega>\frac{2}{3}\left(
1-\lambda\right)  ^{2}$\\
$A_{4}$ & $-\frac{3}{4}\pm\frac{\sqrt{12\omega-7\left(  \lambda-1\right)
^{2}}}{4\left(  \lambda-1\right)  }$ & Attractor\\\hline\hline
\end{tabular}
\label{tab2}%
\end{table}%

We proceed our analysis by considering the exponential scalar field potential.
For arbitrary value of the free parameter $\lambda$ the dynamical system
(\ref{td.09})-(\ref{td.10}) admits the following stationary points~$P=P\left(
x,y\right)  $:

Point $A_{1}$ with coordinates $A_{1}=\left(  0,0\right)  $, in which
$\Omega_{m}\left(  A_{1}\right)  =1$, and $w_{eff}\left(  A_{1}\right)  =0$.
At the stationary point the scalar field is constant, that is $\phi=\phi_{0}$.
Hence, the point describes a universe dominated by the dust fluid source,
where the scale factor is $a\left(  t\right)  =a_{0}t^{\frac{2}{3}}$. In order
to infer on the stability of the stationary points we linearize equations
(\ref{td.09})-(\ref{td.10}) around the stationary point, and we derive the
eigenvalues of the linearized matrix, they are$\ $calculated $e_{1}\left(
A_{1}\right)  =-\frac{3}{2}~$, $e_{2}\left(  A_{1}\right)  =\frac{3}{2}$.
Hence, because only one of the eigenvalues is negative, the stationary point
is always a saddle point.

\begin{figure}[ptb]
\centering\includegraphics[width=1\textwidth]{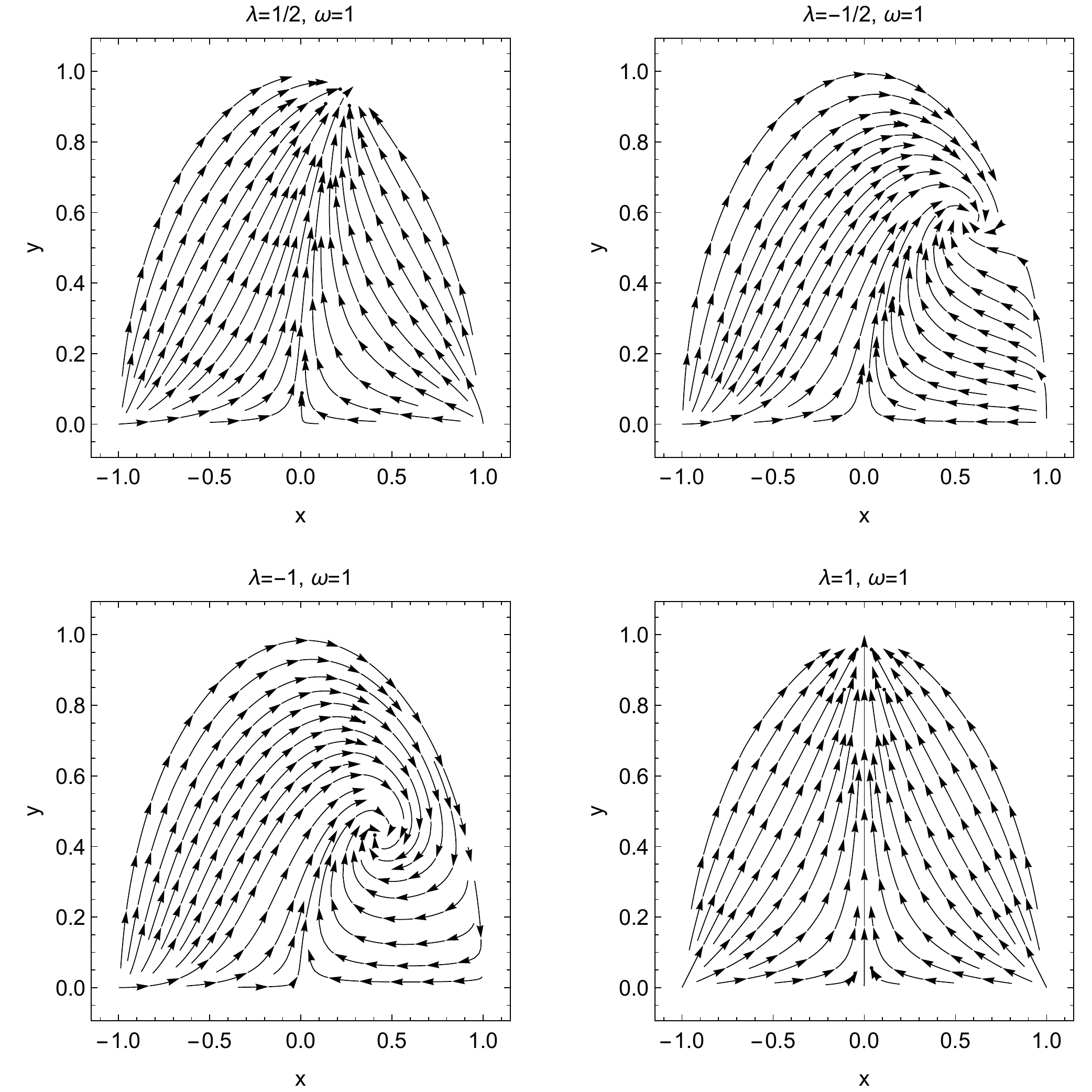}\caption{Two-dimensional
phase-space diagrams for the dynamical system (\ref{td.09})-(\ref{td.10}) in
the plane$~\left(  x,y\right)  $ Left figure in the first row is for $\left(
\lambda,\omega\right)  =\left(  \frac{1}{2},1\right)  $ with attractor point
$A_{3}$. Right figure in the first row is for $\left(  \lambda,\omega\right)
=\left(  -\frac{1}{2},1\right)  $, with attractor point $A_{4}$. For the
figures of the second row, left figure is for $\left(  \lambda,\omega\right)
=\left(  -1,1\right)  $ with attractor point $A_{4}$, while the right figure
is for $\left(  \lambda,\omega\right)  =\left(  1,1\right)  $ and attractor
the de Sitter point $A_{3}$. }%
\label{fig1}%
\end{figure}

The stationary points $A_{2}^{\pm}=\left(  \pm1,0\right)  $ describe universes
where only the scalar field contributes in the cosmological solution, that is,
$\Omega_{m}\left(  A_{2}^{\pm}\right)  =0$. Because $y\left(  A_{2}^{\pm
}\right)  =0$, it means that the scalar field potential does not contribute in
the total cosmological evolution, while only the kinematic part contributes,
that is, $V\left(  \phi\right)  <<\dot{\phi}^{2}$. The effective equation of
state parameter is derived $w_{eff}\left(  A_{2}^{\pm}\right)  =1\mp\frac
{2}{\sqrt{3\omega}}$. For point $A_{2}^{+}$ it follows that $w_{eff}\left(
A_{2}^{+}\right)  <1$. Furthermore, the asymptotic solution at $A_{2}^{+}$
describes an accelerated universe when $\omega>\frac{4}{3}$. On the other
hand, the asymptotic solution at the stationary point $A_{2}^{-}$ gives
$w_{eff}\left(  A_{2}^{-}\right)  >1$. As far as the stability properties of
the points are concerned, we derive the eigenvalues $e_{1}\left(  A_{2}^{\pm
}\right)  =3\pm\sqrt{\frac{3}{\omega}}\left(  \lambda-1\right)  ~$,
$e_{2}\left(  A_{2}^{\pm}\right)  =3$. Thus, point $A_{2}^{+}$ is a saddle
point when $\lambda<1$ and $\omega<\frac{1}{3}\left(  1-\lambda\right)  ^{2}$,
otherwise~$A_{2}^{+}$ is a source. Point $A_{2}^{-}$ a saddle point
when~$\lambda>1$ and $\omega<\frac{1}{3}\left(  1-\lambda^{2}\right)  \,$.

The stationary point $A_{3}=\left(  \frac{1-\lambda}{\sqrt{3\omega}}%
,\sqrt{1-\frac{\left(  1-\lambda\right)  ^{2}}{3\omega}}\right)  ~$is
physically accepted when $\omega>\frac{1}{3}\left(  1-\lambda\right)  ^{2}$.
Point $A_{3}$ describes a scaling solution for an ideal gas with parameter for
the equation of state $w_{eff}\left(  A_{3}\right)  =-1+\frac{2\lambda
}{3\omega}\left(  \lambda-1\right)  $, and $\Omega_{m}\left(  A_{3}\right)
=0$. The point describes an accelerated universe when the free parameters are
constraint as $\left\{  \lambda\leq0,\omega>\lambda\left(  \lambda-1\right)
\right\}  $, $\left\{  0<\lambda\leq1\right\}  $ and $\left\{  \lambda
>1,\omega>\lambda\left(  \lambda-1\right)  \right\}  $. For $\lambda=0$ or
$\lambda=1$, the stationary points describe the de Sitter universe\ i.e.
$w_{eff}\left(  A_{3}\right)  =-1$. The eigenvalues of the linearized system
are derived to be $e_{1}\left(  A_{3}\right)  =-3+\frac{2}{\omega}\left(
1-\lambda\right)  ^{2}~$, $e_{2}\left(  A_{3}\right)  =-3+\frac{\left(
1-\lambda\right)  ^{2}}{\omega}$. Therefore, point $A_{3}$ is an attractor
when $\omega>\frac{2}{3}\left(  1-\lambda\right)  ^{2}$ and for arbitrary
value for the parameter $\lambda$.

Finally, the stationary points $A_{4}=\left(  \frac{\sqrt{3}}{2\left(
1-\lambda\right)  },\sqrt{\frac{3\omega}{4\left(  1-\lambda\right)  ^{2}}%
}\right)  $ exist when $\lambda\neq1$. The latter stationary points describe
scaling solutions with $w_{eff}\left(  A_{4}\right)  =\frac{1}{\lambda-1},$
$\ $for $\lambda\neq0,$ or de Sitter universes $w_{eff}\left(  A_{4}\right)
=-1$ when $\lambda=0.$ In addition, the contribution of the dust fluid source
is nonzero, i.e. $\Omega_{m}\left(  A_{4}\right)  =1-\frac{3\omega}{2\left(
1-\lambda\right)  ^{2}}$. The point is physically accepted, when $0<\Omega
_{m}\left(  A_{4}\right)  <1$, that is, $\lambda\neq1$, $0<\omega<\frac{2}%
{3}\left(  1-\lambda\right)  ^{2}$. The eigenvalues of the linearized system
are derived $e_{1}\left(  A_{4}\right)  =-\frac{3}{4}+\frac{\sqrt
{12\omega-7\left(  \lambda-1\right)  ^{2}}}{4\left(  \lambda-1\right)  }$,
$e_{2}\left(  A_{4}\right)  =-\frac{3}{4}-\frac{\sqrt{12\omega-7\left(
\lambda-1\right)  ^{2}}}{4\left(  \lambda-1\right)  }$. We infer that
point$~A_{4}$ is always an attractor.

\begin{figure}[ptb]
\centering\includegraphics[width=1\textwidth]{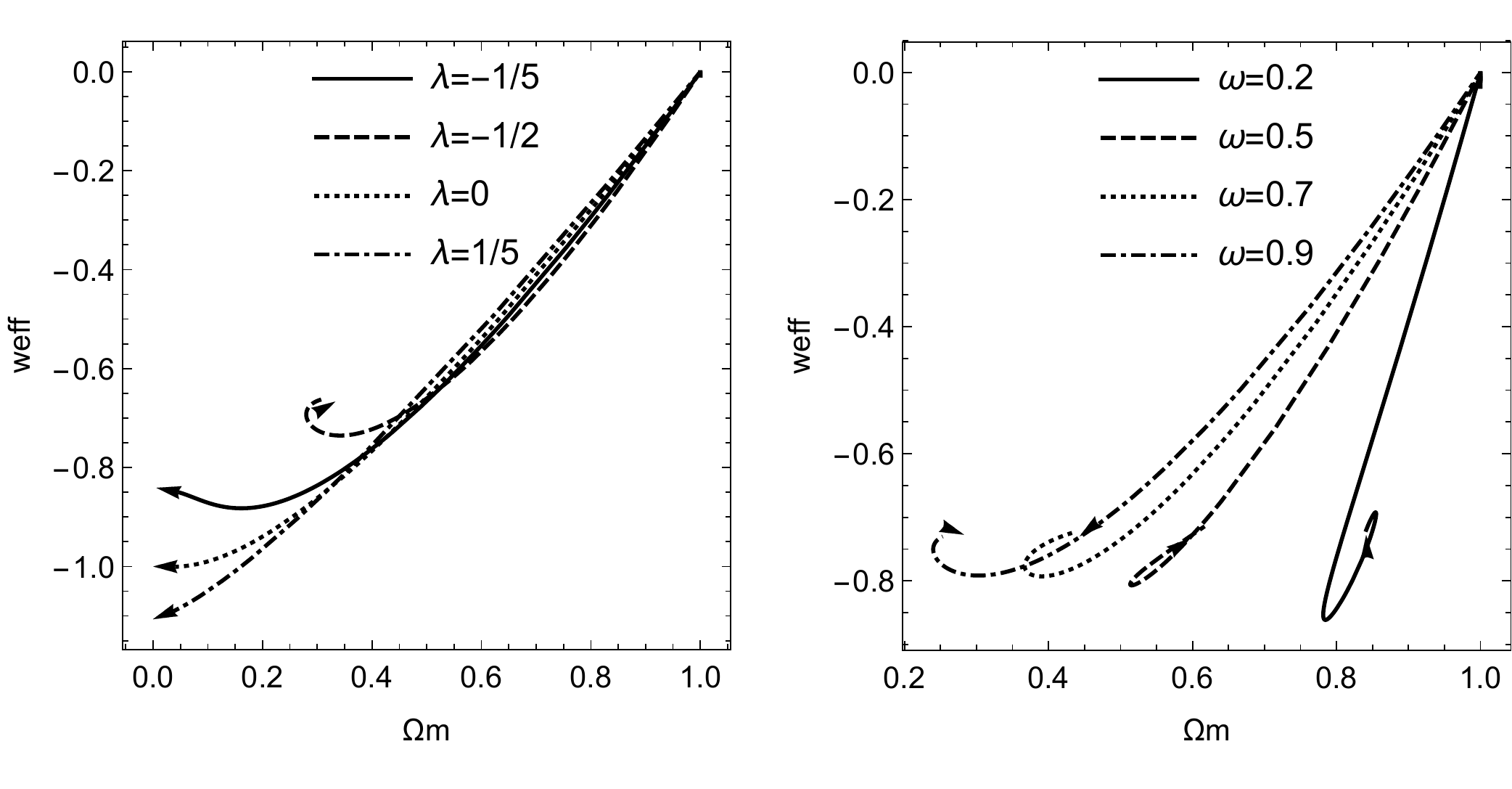}\caption{Parametric
plot for the qualitative evolution of the $\Omega_{m}$ and of the effective
equation of state parameter $w_{eff}$ for numerical solution of the dynamical
system (\ref{td.08})-(\ref{td.10}) with initial conditions near to the matter
dominated era $\Omega_{m}\simeq1$. In Left Fig., lines are for $\omega=1$,
solid line is for $\lambda=-\frac{1}{5}$, dashed line is for $\lambda
=-\frac{1}{2}$, dotted line is for $\lambda=0$ and dashed dot line is for
$\lambda=\frac{1}{5}$. In the Right Fig. lines are for $\lambda=0$, solid line
is for $\omega=0.2$, dashed line is for $\omega=0.5$, dotted line is for
$\omega=0.7$ and dashed dot line is for $\omega=0.9$. $\ $The attractors is
the de Sitter point $A_{4}$. }%
\label{fig22}%
\end{figure}

The results for the stationary points and their physical properties are
summarized in Table \ref{tab1}, while in Table \ref{tab2} we summarize the
stability conditions for the stationary points. In Fig. \ref{fig1} we present
the phase-space portrait for the field equations in the plane $\left(
x,y\right)  $ for different values of the free parameters in which
point~$A_{3}$ or $A_{4}$ are attractors.\ In Figs. \ref{fig22} we present the
parametric plot with the evolution of the physical variables $\Omega_{m}$ and
$w_{eff}$ for numerical solutions of the field equations.

The stationary point $A_{4}$ is of special interest, because it can describe
an accelerated universe where the dark matter and the dark energy contributes
in the cosmological fluid. This point is always an attractor when it exists.
Moreover, from the recent cosmological observations we know that the
deceleration parameter $q=\frac{1}{2}\left(  1+3w_{eff}\right)  $ for the
$\Lambda$-cosmology is $q_{\Lambda}\simeq-0.6$ where $\Omega_{m}\simeq
0.27$~\cite{planck1}. Thus, the effective equation of state parameter is
$w_{eff}^{\Lambda}\simeq-0.73$. Therefore, for$~\left(  \omega,\lambda\right)
\simeq\left(  0.91,-0.37\right)  $ the solution of point $A_{4}$ provide
physical variables equal with that of the $\Lambda$-cosmology.

\section{Cosmographic parameters}

\label{sec4}

The cosmographic approach is a model independent construction way of the
cosmological physical variables \cite{Weinberg-GR}.\ Specifically, the scale
factor it is written in the expansion form%
\begin{equation}
\frac{a(t)}{a_{0}}=1+H_{0}(t-t_{0})-\frac{1}{2}q_{0}^{2}(t-t_{0})^{2}+\frac
{1}{3!}j_{0}(t-t_{0})^{3}+\frac{1}{4!}s_{0}(t-t_{0})^{4}+\mathcal{O}%
[(t-t_{0})^{5}]~,
\end{equation}
where $H_{0}$ is the value of the Hubble function of today, $q_{0}$ is the
deceleration parameter of today, $j_{0}$ and $s_{0}$ are the present value for
the jerk and snap parameters. The $H$, $q$, $j$, $s$ are kinematical
quantities which are extracted directly from the spacetime. The kinematic
quantities are defined as \cite{Visser:2003vq,Bolotin:2018xtq}
\begin{align}
&  H(t)=\frac{1}{a}\frac{da}{dt},\nonumber\\
&  q(t)=-\frac{1}{a}\frac{d^{2}a}{dt^{2}}\left[  \frac{1}{a}\frac{da}%
{dt}\right]  ^{-2},\nonumber\\
&  j(t)=\frac{1}{a}\frac{d^{3}a}{dt^{3}}\left[  \frac{1}{a}\frac{da}%
{dt}\right]  ^{-3},\nonumber\\
&  s(t)=\frac{1}{a}\frac{d^{4}a}{dt^{4}}\left[  \frac{1}{a}\frac{da}%
{dt}\right]  ^{-4},\nonumber
\end{align}
or in terms of the Hubble function they can be written in the equivalent form
\begin{align}
q  &  =-1-\frac{\dot{H}}{H^{2}}\\
j  &  =\frac{\ddot{H}}{H^{3}}-3q-2\\
s  &  =\frac{\dddot{H}}{H^{4}}+4j+3q\left(  q+4\right)  +6,
\end{align}

\begin{figure}[ptb]
\centering\includegraphics[width=1\textwidth]{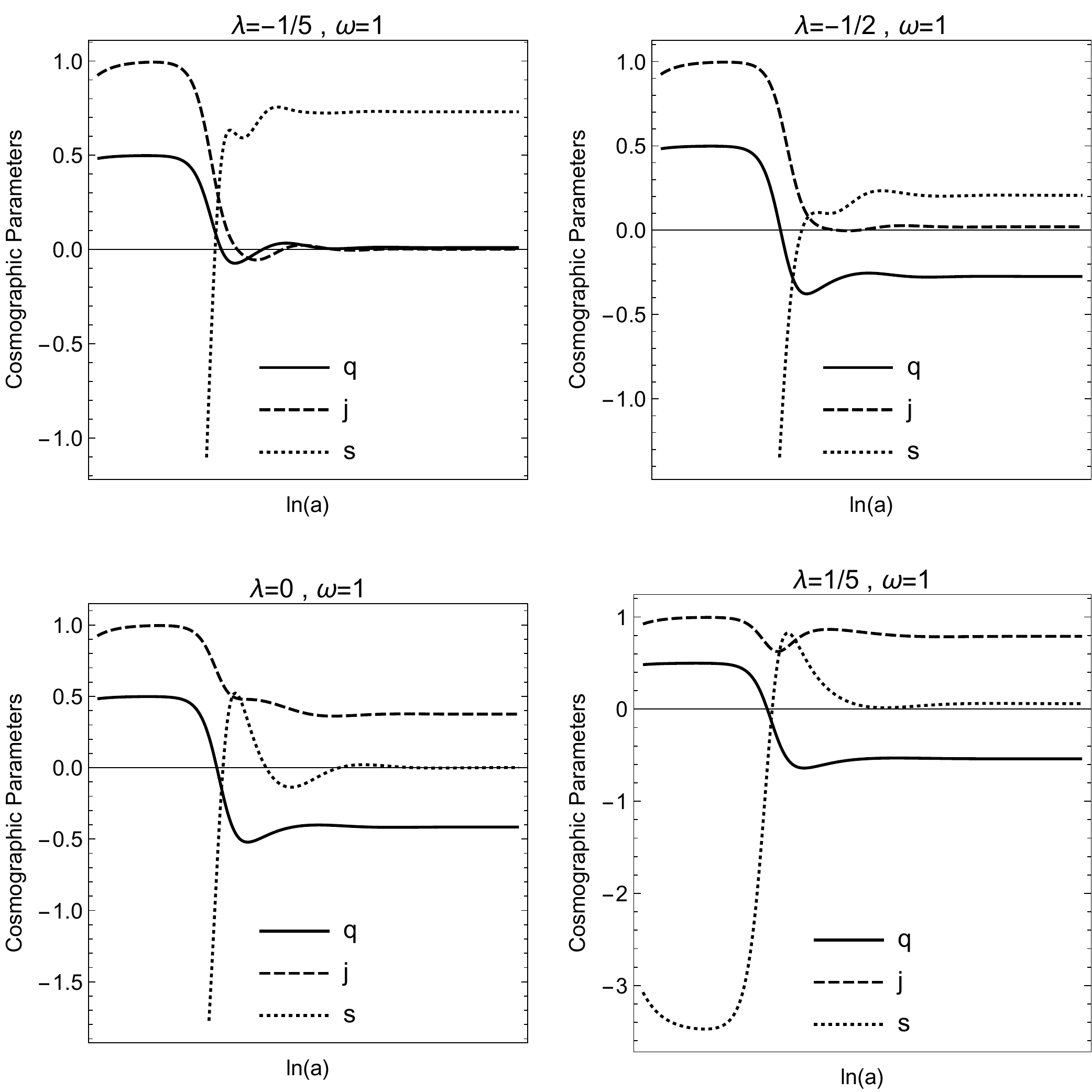}\caption{Qualitative
evolution for the cosmographic parameters $q$, $j$, and $s$ as provided by the
field equations (\ref{td.09})-(\ref{td.10}) for different values of the free
parameters. The initial condition is a point near to the matter dominated
solution $A_{1}$. Solid line is for decellaration parameter $q$, dashed line
is for the jerk parameter $j$ and dotted line is for the snap parameter $s$. }%
\label{fig3}%
\end{figure}

\begin{figure}[ptb]
\centering\includegraphics[width=1\textwidth]{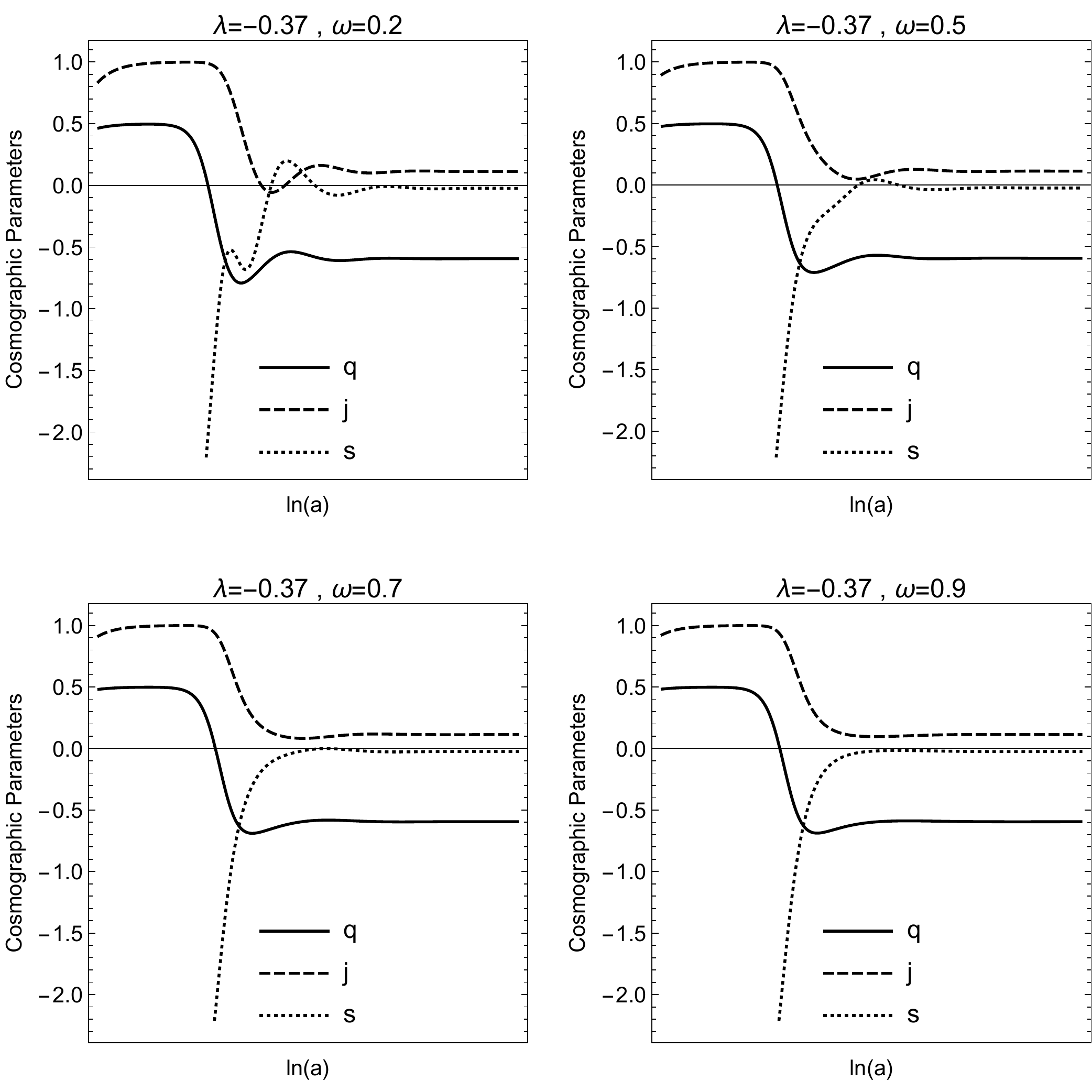}\caption{Qualitative
evolution for the cosmographic parameters $q$, $j$, and $s$ as provided by the
field equations (\ref{td.09})-(\ref{td.10}) for different values of the free
parameters. The initial condition is a point near to the matter dominated
solution $A_{1}$. Solid line is for decellaration parameter $q$, dashed line
is for the jerk parameter $j$ and dotted line is for the snap parameter $s$. }%
\label{fig4}%
\end{figure}

Thus, from the evolution of the cosmographic $q,j,s$, one can understand the
expansion of the universe, the rate of acceleration and its derivative. In
Figs. \ref{fig3} and \ref{fig4} we present the qualitative evolution of the
cosmographic parameters for the model of our consideration for initial
conditions near to the matter dominated era and for the values of the free
parameters $\left(  \lambda,\omega\right)  $ same with that of the numerical
solutions of Fig. \ref{fig22}. From the evolution of the cosmographic
parameters presented in Figs. \ref{fig3} and \ref{fig4} we observe that while
the $q_{0}$ value for the $\Lambda$CDM is recovered that it is not true for
the jerk and snap parameters which in general have different evolution.

From the qualitative evolution we observe that our model can predict values
for the cosmographic parameters as they given by the cosmological constraints
\cite{cosmp}.

\section{Conclusions}

\label{sec5}

In this study we considered a scalar-torsion model known as teleparallel
dilaton theory coupled to a pressureless fluid source, which we assume that it
describes the dark matter. For this cosmological model the field equations are
of second order and we investigated the evolution of the cosmological
parameters in a spatially flat FLRW background space by determine the
stationary points and study their stability.

This kind of dynamical analysis is essential for the study of the general
evolution of the dynamical system because it provide us with important
information in order to infer about the cosmological viability of theory. For
the model of our consideration we wrote the field equations into an equivalent
algebraic-differential by using the $H-$normalization approach. Every
stationary point of the dynamical system describes an exact asymptotic
solution for the scale factor which corresponds to a specific epoch of the
cosmological history. The stability properties of the stationary points are
important to investigated because they tell us about the general evolution of
the dynamical system. \newline

For our model, and for the exponential scalar field potential, we found that
the field equations admit four stationary points which describe four different
cosmological epochs. Point $A_{1}$ provides the exact solution of the unstable
matter dominated era; points $A_{2}^{\pm}$ describe universes dominated by the
kinetic part of the scalar field where in contrary to the quintessence at this
theory we calculate $w_{eff}\left(  A_{1}^{\pm}\right)  =1\mp\frac{2}%
{\sqrt{3\omega}}$. Hence for $\omega>\frac{4}{3}$ point $A_{2}^{+}$ describes
the cosmic acceleration. Point $A_{3}$ is a scaling solution in general
$w_{eff}\left(  A_{3}\right)  =-1+\frac{2\lambda\left(  \lambda-1\right)
}{3\omega}$, where for $\lambda\left(  \lambda-1\right)  =0,$ the asymptotic
solution at point $A_{3}$ is the de Sitter solution. Finally, point $A_{4}$
describes an asymptotic solution in which the dark matter and the scalar field
contributes in the cosmological fluid.\ This solution is of special interest
because it describes the present cosmological era. Moreover, in order to
understand the evolution of the physical variables we presented the
qualitative evolution of the cosmographic parameters from where we found that
the values for some of the cosmographic parameters at the present era can be recovered.

From the above results it is clear that while there are similarities of this
model with the dilaton model of scalar tensor theory \cite{f1}, the two
theories are different in the general evolution of the cosmological history.
In the same conclusion we end by comparing the results of this work that
previous studies on the Weyl Integrable theory, where interaction of the
scalar field with the dust fluid is introduced in the gravitational Action
Integral \cite{f2}. Consequently, the present model of study has interesting
properties which can explain the cosmic history and deserves further study. In
addition, in the absence of the dust fluid we recall that the theory admits a
discrete symmetry which can be used to study the cosmic evolution in the
pre-big bang era as in string cosmology. However, in contrary to string
cosmology and the dilaton field for this teleparallel model the pre-big bang
era is not recovered as reflection of the present epoch.

In a future study, we plan to use the cosmological observations in order to
constraint this specific theory as a dark energy candidate.

\end{document}